# The evolution of the Earth-Moon system based on the dark matter field fluid model

Hongjun Pan

Department of Chemistry

University of North Texas, Denton, Texas 76203, U. S. A.

**Abstract**

The evolution of Earth-Moon system is described by the dark matter field fluid model with a non-Newtonian approach proposed in the Meeting of Division of Particle and Field 2004, American Physical Society. The current behavior of the Earth-Moon system agrees with this model very well and the general pattern of the evolution of the Moon-Earth system described by this model agrees with geological and fossil evidence. The closest distance of the Moon to Earth was about 259000 *km* at 4.5 billion years ago, which is far beyond the Roche's limit. The result suggests that the tidal friction may not be the primary cause for the evolution of the Earth-Moon system. The average dark matter field fluid constant derived from Earth-Moon system data is 4.39 × $10^{-22}$ $s^{-1}m^{-1}$. This model predicts that the Mars's rotation is also slowing with the angular acceleration rate about -4.38 × $10^{-22}$ *rad* $s^{-2}$.

# 1. Introduction

The popularly accepted theory for the formation of the Earth-Moon system is that the Moon was formed from debris of a strong impact by a giant planetesimal with the Earth at the close of the planet-forming period (Hartmann and Davis 1975). Since the formation of the Earth-Moon system, it has been evolving at all time scale. It is well known that the Moon is receding from us and both the Earth's rotation and Moon's rotation are slowing. The popular theory is that the tidal friction causes all those changes based on the conservation of the angular momentum of the Earth-Moon system. The situation becomes complicated in describing the past evolution of the Earth-Moon system. Because the Moon is moving away from us and the Earth rotation is slowing, this means that the Moon was closer and the Earth rotation was faster in the past. Creationists argue that based on the tidal friction theory, the tidal friction should be stronger and the recessional rate of the Moon should be greater in the past, the distance of the Moon would quickly fall inside the Roche's limit (for earth, 15500 *km*) in which the Moon would be torn apart by gravity in 1 to 2 billion years ago. However, geological evidence indicates that the recession of the Moon in the past was slower than the present rate, i. e., the recession has been accelerating with time. Therefore, it must be concluded that tidal friction was very much less in the remote past than we would deduce on the basis of present-day observations (Stacey 1977). This was called "geological time scale difficulty" or "Lunar crisis" and is one of the main arguments by creationists against the tidal friction theory (Brush 1983).

But we have to consider the case carefully in various aspects. One possible scenario is that the Earth has been undergoing dynamic evolution at all time scale since its inception, the geological and physical conditions (such as the continent positions and drifting, the crust, surface temperature fluctuation like the glacial/snowball effect, etc) at remote past could be substantially different from currently, in which the tidal friction could be much less; therefore, the receding rate of the Moon could be slower. Various tidal friction models were proposed in the past to describe the evolution of the Earth-Moon system to avoid such difficulty or crisis and put the Moon at quite a comfortable distance from Earth at 4.5 billion years ago (Hansen 1982, Kagan and Maslova 1994, Ray et al. 1999, Finch 1981, Slichter 1963). The tidal friction theories explain that the present rate of tidal dissipation is anomalously high because the tidal force is close to a resonance in the response function of ocean (Brush 1983). Kagan gave a detailed review about those tidal friction models (Kagan 1997). Those models are based on many assumptions about geological (continental position and drifting) and physical conditions in the past, and many parameters (such as phase lag angle, multi-mode approximation with time dependent frequencies of the resonance modes, etc.) have to be introduced and carefully adjusted to make their predictions close to the geological evidence. However, those assumptions and parameters are still challenged, to certain extent, as concoction.

The second possible scenario is that another mechanism could dominate the evolution of the Earth-Moon system and the role of the tidal friction is not significant. In the Meeting of Division of Particle and Field 2004, American Physical Society, University of California at Riverside, the author proposed a dark matter field fluid model (Pan 2005) with a non-Newtonian approach, the current Moon and Earth data agree with this model very well. This paper will demonstrate that the past evolution of Moon-Earth system can be described by the dark matter field fluid model without any assumptions about past geological and physical conditions. Although the subject of the evolution of the Earth-Moon system has been extensively studied analytically or numerically, to the author's knowledge, there are no theories similar or equivalent to this model.

## 2. Invisible matter

In modern cosmology, it was proposed that the visible matter in the universe is about 2 ~ 10 % of the total matter and about 90 ~ 98% of total matter is currently invisible which is called dark matter and dark energy, such invisible matter has an anti-gravity property to make the universe expanding faster and faster.

If the ratio of the matter components of the universe is close to this hypothesis, then, the evolution of the universe should be dominated by the physical mechanism of such invisible matter, such physical mechanism could be far beyond the current Newtonian physics and Einsteinian physics, and the Newtonian physics and Einsteinian physics could reflect only a corner of the iceberg of the greater physics.

If the ratio of the matter components of the universe is close to this hypothesis, then, it should be more reasonable to think that such dominant invisible matter spreads in everywhere of the universe (the density of the invisible matter may vary from place to place); in other words, all visible matter objects should be surrounded by such invisible matter and the motion of the visible matter objects should be affected by the invisible matter if there are interactions between the visible matter and the invisible matter.

If the ratio of the matter components of the universe is close to this hypothesis, then, the size of the particles of the invisible matter should be very small and below the detection limit of the current technology; otherwise, it would be detected long time ago with such dominant amount.

With such invisible matter in mind, we move to the next section to develop the dark matter field fluid model with non-Newtonian approach. For simplicity, all invisible matter (dark matter, dark energy and possible other terms) is called dark matter here.

### 3. The dark matter field fluid model

In this proposed model, it is assumed that:

1. A celestial body rotates and moves in the space, which, for simplicity, is uniformly filled with the dark matter which is in quiescent state relative to the motion of the celestial body. The dark matter possesses a field property and a fluid property; it can interact with the celestial body with its fluid and field properties; therefore, it can have energy exchange with the celestial body, and affect the motion of the celestial body.

2. The fluid property follows the general principle of fluid mechanics. The dark matter field fluid particles may be so small that they can easily permeate into ordinary "baryonic" matter; i. e., ordinary matter objects could be saturated with such dark matter field fluid. Thus, the whole celestial body interacts with the dark matter field fluid, in the manner of a sponge moving thru water. The nature of the field property of the dark matter field fluid is unknown. It is here assumed that the interaction of the field associated with the dark matter field fluid with the celestial body is proportional to the mass of the celestial body. The dark matter field fluid is assumed to have a repulsive force against the gravitational force towards baryonic matter. The nature and mechanism of such repulsive force is unknown.

With the assumptions above, one can study how the dark matter field fluid may influence the motion of a celestial body and compare the results with observations. The common shape of celestial bodies is spherical. According to Stokes's law, a rigid non-permeable sphere moving through a quiescent fluid with a sufficiently low Reynolds number experiences a resistance force $F$

$$F = -6\pi\mu r v \tag{1}$$

where $v$ is the moving velocity, $r$ is the radius of the sphere, and $\mu$ is the fluid viscosity constant. The direction of the resistance force $F$ in Eq. 1 is opposite to the direction of the velocity $v$. For a rigid sphere moving through the dark matter field fluid, due to the dual properties of the dark matter field fluid and its permeation into the sphere, the force $F$ may not be proportional to the radius of the sphere. Also, $F$ may be proportional to the mass of the sphere due to the field interaction. Therefore, with the combined effects of both fluid and field, the force exerted on the sphere by the dark matter field fluid is assumed to be of the scaled form

$$F = -6\pi\eta r^{1-n} m v \tag{2}$$

where $n$ is a parameter arising from saturation by dark matter field fluid, the $r^{1-n}$ can be viewed as the effective radius with the same unit as $r$, $m$ is the mass of the sphere, and $\eta$ is the dark matter field fluid constant, which is equivalent to $\mu$. The direction of the resistance force $F$ in Eq. 2 is opposite to the direction of the velocity $v$. The force described by Eq. 2 is velocity-dependent and causes negative acceleration. According to Newton's second law of motion, the equation of motion for the sphere is

$$m\frac{dv}{dt} = -6\pi\eta r^{1-n} m v \tag{3}$$

Then

$$v = v_0 \exp(-6\pi\eta r^{1-n} v t) \tag{4}$$

where $v_0$ is the initial velocity ($t = 0$) of the sphere. If the sphere revolves around a massive gravitational center, there are three forces in the line between the sphere and the gravitational center: (1) the gravitational force, (2) the centripetal acceleration force; and (3) the repulsive force of the dark matter field fluid. The drag force in Eq. 3 reduces the orbital velocity and causes the sphere to move inward to the gravitational center. However, if the sum of the centripetal acceleration force and the repulsive force is stronger than the gravitational force, then, the sphere will move outward and recede from the gravitational center. This is the case of interest here. If the velocity change in Eq. 3 is sufficiently slow and the repulsive force is small compared to the gravitational force and centripetal acceleration force, then the rate of receding will be accordingly relatively slow. Therefore, the gravitational force and the centripetal acceleration force can be approximately treated in equilibrium at any time. The pseudo equilibrium equation is

$$\frac{GMm}{R^2} = \frac{mv^2}{R} \tag{5}$$

where $G$ is the gravitational constant, $M$ is the mass of the gravitational center, and $R$ is the radius of the orbit. Inserting $v$ of Eq. 4 into Eq. 5 yields

$$R = \frac{GM}{v_0^2} \exp(12\pi\eta r^{1-n} t) \tag{6}$$

or

$$R = R_0 \exp(12\pi\eta r^{1-n} t) \tag{7}$$

where

$$R_0 = \frac{GM}{v_0^2} \tag{8}$$

$R_0$ is the initial distance to the gravitational center. Note that $R$ exponentially increases with time. The increase of orbital energy with the receding comes from the repulsive force of dark matter field fluid. The recessional rate of the sphere is

$$\frac{dR}{dt} = 12\pi\eta r^{1-n} R \tag{9}$$

The acceleration of the recession is

$$\frac{d^2 R}{dt^2} = \left(12\pi\eta r^{1-n}\right)^2 R. \tag{10}$$

The recessional acceleration is positive and proportional to its distance to the gravitational center, so the recession is faster and faster.

According to the mechanics of fluids, for a rigid non-permeable sphere rotating about its central axis in the quiescent fluid, the torque $T$ exerted by the fluid on the sphere is

$$T = -8\pi\mu r^3 \omega \tag{11}$$

where $\omega$ is the angular velocity of the sphere. The direction of the torque in Eq. 11 is opposite to the direction of the rotation. In the case of a sphere rotating in the quiescent dark matter field fluid with angular velocity $\omega$, similar to Eq. 2, the proposed $T$ exerted on the sphere is

$$T = -8\pi\eta \left(r^{1-n}\right)^3 m\omega \tag{12}$$

The direction of the torque in Eq. 12 is opposite to the direction of the rotation. The torque causes the negative angular acceleration

$$T = I\frac{d\omega}{dt} \tag{13}$$

where $I$ is the moment of inertia of the sphere in the dark matter field fluid

$$I = \frac{2}{5}m\left(r^{1-n}\right)^2 \tag{14}$$

Therefore, the equation of rotation for the sphere in the dark matter field fluid is

$$\frac{d\omega}{dt} = -20\pi\eta r^{1-n}\omega \tag{15}$$

Solving this equation yields

$$\omega = \omega_0 \exp(-20\pi\eta r^{1-n}t) \tag{16}$$

where $\omega_0$ is the initial angular velocity. One can see that the angular velocity of the sphere exponentially decreases with time and the angular deceleration is proportional to its angular velocity.

For the same celestial sphere, combining Eq. 9 and Eq. 15 yields

$$\frac{\frac{1}{\omega}\frac{d\omega}{dt}}{\frac{1}{R}\frac{dR}{dt}} = -\frac{5}{3} = -1.67 \tag{17}$$

The significance of Eq. 17 is that it contains only observed data without assumptions and undetermined parameters; therefore, it is a critical test for this model.

For two different celestial spheres in the same system, combining Eq. 9 and Eq. 15 yields

$$\frac{\frac{1}{\omega_1}\frac{d\omega_1}{dt}}{\frac{1}{R_2}\frac{dR_2}{dt}}\left(\frac{r_2}{r_1}\right)^{1-n} = -\frac{5}{3} = -1.67 \tag{18}$$

This is another critical test for this model.

## 4. The current behavior of the Earth-Moon system agrees with the model

The Moon-Earth system is the simplest gravitational system. The solar system is complex, the Earth and the Moon experience not only the interaction of the Sun but also interactions of other planets. Let us consider the local Earth-Moon gravitational system as an isolated local gravitational system, i.e., the influence from the Sun and other planets on the rotation and orbital motion of the Moon and on the rotation of the Earth is assumed negligible compared to the forces exerted by the moon and earth on each other. In addition, the eccentricity of the Moon's orbit is small enough to be ignored. The data about the Moon and the Earth from references (Dickey et .al., 1994, and Lang, 1992) are listed below for the readers' convenience to verify the calculation because the data may vary slightly with different data sources.

Moon:

Mean radius:  $r = 1738.0\ km$

Mass:  $m = 7.3483 \times 10^{25}\ gram$

Rotation period = $27.321661\ days$

Angular velocity of Moon = $2.6617 \times 10^{-6}\ rad\ s^{-1}$

Mean distance to Earth $R_m = 384400\ km$

Mean orbital velocity $v = 1.023\ km\ s^{-1}$

Orbit eccentricity  $e = 0.0549$

Angular rotation acceleration rate = $-25.88 \pm 0.5\ arcsec\ century^{-2}$

$= (-1.255 \pm 0.024) \times 10^{-4}\ rad\ century^{-2}$

$= (-1.260 \pm 0.024) \times 10^{-23}\ rad\ s^{-2}$

Receding rate from Earth = $3.82 \pm 0.07\ cm\ year^{-1} = (1.21 \pm 0.02) \times 10^{-9}\ m\ s^{-1}$

Earth:

   Mean radius: $r = 6371.0 \, km$

   Mass: $m = 5.9742 \times 10^{27} \, gram$

   Rotation period = 23 h 56m 04.098904s = 86164.098904s

   Angular velocity of rotation = $7.292115 \times 10^{-5} \, rad \, s^{-1}$

   Mean distance to the Sun $R_m$= 149,597,870.61 $km$

   Mean orbital velocity $v = 29.78 \, km \, s^{-1}$

   Angular acceleration of Earth = $(-5.5 \pm 0.5) \times 10^{-22} \, rad \, s^{-2}$

The Moon's angular rotation acceleration rate and increase in mean distance to the Earth (receding rate) were obtained from the lunar laser ranging of the Apollo Program (Dickey et .al., 1994). By inserting the data of the Moon's rotation and recession into Eq. 17, the result is

$$\frac{-1.26 \times 10^{-23} \times 3.92509 \times 10^{8}}{1.21 \times 10^{-9} \times 2.662 \times 10^{-6}} = -1.54 \pm 0.039 \quad (19)$$

The distance R in Eq. 19 is from the Moon's center to the Earth's center and the number 384400 $km$ is assumed to be the distance from the Moon's surface to the Earth's surface. Eq. 19 is in good agreement with the theoretical value of -1.67. The result is in accord with the model used here. The difference (about 7.8%) between the values of -1.54 and -1.67 may come from several sources:

    1. Moon's orbital is not a perfect circle

    2. Moon is not a perfect rigid sphere.

    3. The effect from Sun and other planets.

    4. Errors in data.

    5. Possible other unknown reasons.

The two parameters *n* and *η* in Eq. 9 and Eq. 15 can be determined with two data sets. The third data set can be used to further test the model. If this model correctly describes the situation at hand, it should give consistent results for different motions. The values of *n* and *η* calculated from three different data sets are listed below (Note, the mean distance of the Moon to the Earth and mean radii of the Moon and the Earth are used in the calculation).

The value of *n*: $\quad n = 0.64$

From the Moon's rotation: $\eta = 4.27 \times 10^{-22} \ s^{-1} \ m^{-1}$

From the Earth's rotation: $\eta = 4.26 \times 10^{-22} \ s^{-1} \ m^{-1}$

From the Moon's recession: $\eta = 4.64 \times 10^{-22} \ s^{-1} \ m^{-1}$

One can see that the three values of *η* are consistent within the range of error in the data.

The average value of *η*: $\eta = (4.39 \pm 0.22) \times 10^{-22} \ s^{-1} \ m^{-1}$

By inserting the data of the Earth's rotation, the Moon's recession and the value of *n* into Eq. 18, the result is

$$\frac{-5.5 \times 10^{-22} \times 3.92509 \times 10^{8}}{7.29 \times 10^{-5} \times 1.21 \times 10^{-9}} \left(\frac{1738000}{6371000}\right)^{(1-0.64)} = -1.53 \pm 0.14 \quad (20)$$

This is also in accord with the model used here.

The dragging force exerted on the Moon's orbital motion by the dark matter field fluid is $-1.11 \times 10^{8}$ *N*, this is negligibly small compared to the gravitational force between the Moon and the Earth ~ $1.90 \times 10^{20}$ *N*; and the torque exerted by the dark matter field fluid on the Earth's and the Moon's rotations is T = $-5.49 \times 10^{16}$ *Nm* and $-1.15 \times 10^{12}$ *Nm*, respectively.

## 5. The evolution of Earth-Moon system

Sonett *et al.* found that the length of the terrestrial day 900 million years ago was about 19.2 hours based on the laminated tidal sediments on the Earth (Sonett *et al.*, 1996). According to the model presented here, back in that time, the length of the day was about 19.2 hours, this agrees very well with Sonett *et al.*'s result.

Another critical aspect of modeling the evolution of the Earth-Moon system is to give a reasonable estimate of the closest distance of the Moon to the Earth when the system was established at 4.5 billion years ago. Based on the dark matter field fluid model, and the above result, the closest distance of the Moon to the Earth was about 259000 *km* (center to center) or 250900 *km* (surface to surface) at 4.5 billion years ago, this is far beyond the Roche's limit. In the modern astronomy textbook by Chaisson and McMillan (Chaisson and McMillan, 1993, p.173), the estimated distance at 4.5 billion years ago was 250000 *km*, this is probably the most reasonable number that most astronomers believe and it agrees excellently with the result of this model. The closest distance of the Moon to the Earth by Hansen's models was about 38 Earth radii or 242000 *km* (Hansen, 1982).

According to this model, the length of day of the Earth was about 8 hours at 4.5 billion years ago. Fig. 1 shows the evolution of the distance of Moon to the Earth and the length of day of the Earth with the age of the Earth-Moon system described by this model along with data from Kvale *et al*. (1999), Sonett *et al*. (1996) and Scrutton (1978). One can see that those data fit this model very well in their time range.

Fig. 2 shows the geological data of solar days year$^{-1}$ from Wells (1963) and from Sonett et al. (1996) and the description (solid line) by this dark matter field fluid model for past 900 million years. One can see that the model agrees with the geological and fossil data beautifully.

The important difference of this model with early models in describing the early evolution of the Earth-Moon system is that this model is based only on current data of the Moon-Earth system and there are no assumptions about the conditions of earlier Earth rotation and continental drifting. Based on this model, the Earth-Moon system has been smoothly evolving to the current position since it was established and the recessional rate of the Moon has been gradually increasing, however, this description does not take it into account that there might be special events happened in the past to cause the suddenly significant changes in the motions of the Earth and the Moon, such as strong impacts by giant asteroids and comets, etc, because those impacts are very common in the universe. The general pattern of the evolution of the Moon-Earth system described by this model agrees with geological evidence. Based on Eq. 9, the recessional rate exponentially increases with time. One may then imagine that the recessional rate will quickly become very large. The increase is in fact extremely slow. The Moon's recessional rate will be $3.04 \times 10^{-9}$ $m\ s^{-1}$ after 10 billion years and $7.64 \times 10^{-9}$ $m\ s^{-1}$ after 20 billion years. However, whether the Moon's recession will continue or at some time later another mechanism will take over is not known. It should be understood that the tidal friction does affect the evolution of the Earth itself such as the surface crust structure, continental drifting and evolution of bio-system, etc; it may also play a role in slowing the Earth's rotation, however, such role is not a dominant mechanism.

Unfortunately, there is no data available for the changes of the Earth's orbital motion and all other members of solar system. According to this model and above results, the recessional rate of the Earth should be $6.86 \times 10^{-7}$ *m s$^{-1}$* = 21.6 *m year$^{-1}$* = 2.16 *km century$^{-1}$*, the length of a year increases about 6.8 *ms* and the change of the temperature is $-1.8 \times 10^{-8}$ *K year$^{-1}$* with constant radiation level of the Sun and the stable environment on the Earth. The length of a year at 1 billion years ago would be 80% of the current length of the year. However, much evidence (growth-bands of corals and shellfish as well as some other evidences) suggest that there has been no apparent change in the length of the year over the billion years and the Earth's orbital motion is more stable than its rotation. This suggests that dark matter field fluid is circulating around Sun with the same direction and similar speed of Earth (at least in the Earth's orbital range). Therefore, the Earth's orbital motion experiences very little or no dragging force from the dark matter field fluid. However, this is a conjecture, extensive research has to be conducted to verify if this is the case.

## 6. Skeptical description of the evolution of the Mars

The Moon does not have liquid fluid on its surface, even there is no air, therefore, there is no ocean-like tidal friction force to slow its rotation; however, the rotation of the Moon is still slowing at significant rate of $(-1.260 \pm 0.024) \times 10^{-23}$ *rad s$^{-2}$*, which agrees with the model very well. Based on this, one may reasonably think that the Mars's rotation should be slowing also.

The Mars is our nearest neighbor which has attracted human's great attention since ancient time. The exploration of the Mars has been heating up in recent decades. NASA, Russian and Europe Space Agency sent many space crafts to the Mars to collect data and study this mysterious planet. So far there is still not enough data about the history of this planet to describe its evolution. Same as the Earth, the Mars rotates about its central axis and revolves around the Sun, however, the Mars does not have a massive moon circulating it (Mars has two small satellites: Phobos and Deimos) and there is no liquid fluid on its surface, therefore, there is no apparent ocean-like tidal friction force to slow its rotation by tidal friction theories. Based on the above result and current the Mars's data, this model predicts that the angular acceleration of the Mars should be about $-4.38 \times 10^{-22}$ $rad\ s^{-2}$. Figure 3 describes the possible evolution of the length of day and the solar days/Mars year, the vertical dash line marks the current age of the Mars with assumption that the Mars was formed in a similar time period of the Earth formation. Such description was not given before according to the author's knowledge and is completely skeptical due to lack of reliable data. However, with further expansion of the research and exploration on the Mars, we shall feel confident that the reliable data about the angular rotation acceleration of the Mars will be available in the near future which will provide a vital test for the prediction of this model. There are also other factors which may affect the Mars's rotation rate such as mass redistribution due to season change, winds, possible volcano eruptions and Mars quakes. Therefore, the data has to be carefully analyzed.

## 7. Discussion about the model

From the above results, one can see that the current Earth-Moon data and the geological and fossil data agree with the model very well and the past evolution of the Earth-Moon system can be described by the model without introducing any additional parameters; this model reveals the interesting relationship between the rotation and receding (Eq. 17 and Eq. 18) of the same celestial body or different celestial bodies in the same gravitational system, such relationship is not known before. Such success can not be explained by "coincidence" or "luck" because of so many data involved (current Earth's and Moon's data and geological and fossil data) if one thinks that this is just a "*ad hoc*" or a wrong model, although the chance for the natural happening of such "coincidence" or "luck" could be greater than wining a jackpot lottery; the future Mars's data will clarify this; otherwise, a new theory from different approach can be developed to give the same or better description as this model does. It is certain that this model is not perfect and may have defects, further development may be conducted.

James Clark Maxwell said in the 1873 " *The vast interplanetary and interstellar regions will no longer be regarded as waste places in the universe, which the Creator has not seen fit to fill with the symbols of the manifold order of His kingdom. We shall find them to be already full of this wonderful medium; so full, that no human power can remove it from the smallest portion of space, or produce the slightest flaw in its infinite continuity. It extends unbroken from star to star ....*" The medium that Maxwell talked about is the aether which was proposed as the carrier of light wave propagation. The Michelson-Morley experiment only proved that the light wave propagation does not depend on such medium and did not reject the existence of the medium in the interstellar space. In fact, the concept of the interstellar medium has been developed dramatically recently such as the dark matter, dark energy, cosmic fluid, etc. The dark matter field fluid is just a part of such wonderful medium and "precisely" described by Maxwell.

## 7. Conclusion

The evolution of the Earth-Moon system can be described by the dark matter field fluid model with non-Newtonian approach and the current data of the Earth and the Moon fits this model very well. At 4.5 billion years ago, the closest distance of the Moon to the Earth could be about 259000 *km*, which is far beyond the Roche's limit and the length of day was about 8 hours. The general pattern of the evolution of the Moon-Earth system described by this model agrees with geological and fossil evidence. The tidal friction may not be the primary cause for the evolution of the Earth-Moon system. The Mars's rotation is also slowing with the angular acceleration rate about $-4.38 \times 10^{-22}$ *rad s$^{-2}$*.

**Caption**

Figure 1, the evolution of Moon's distance and the length of day of the earth with the age of the Earth-Moon system. Solid lines are calculated according to the dark matter field fluid model. Data sources: the Moon distances are from Kvale and *et a*l. and for the length of day: (a and b) are from Scrutton ( page 186, fig. 8), c is from Sonett and *et al*. The dash line marks the current age of the Earth-Moon system.

Figure 2, the evolution of Solar days of year with the age of the Earth-Moon system. The solid line is calculated according to dark matter field fluid model. The data are from Wells (3.9 ~ 4.435 billion years range), Sonett (3.6 billion years) and current age (4.5 billion years).

Figure 3, the skeptical description of the evolution of Mars's length of day and the solar days/Mars year with the age of the Mars (assuming that the Mars's age is about 4.5 billion years). The vertical dash line marks the current age of Mars.

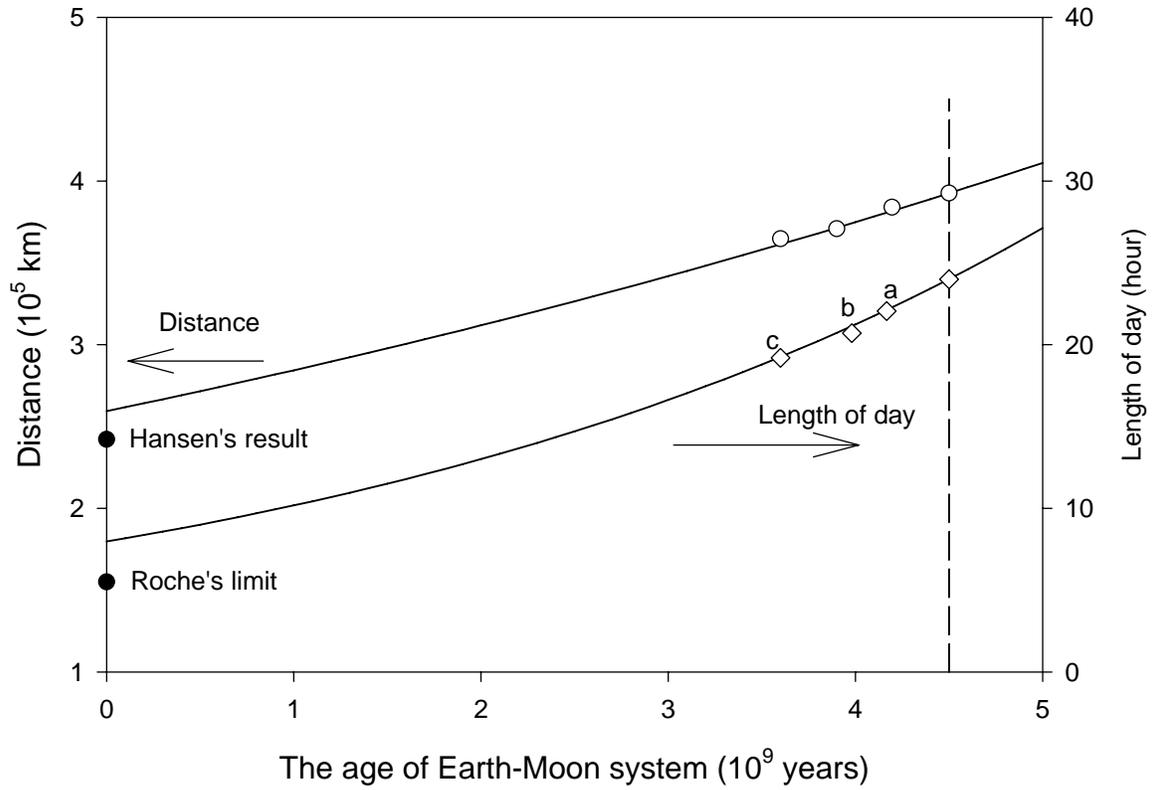

Figure 1, Moon's distance and the length of day of Earth change with the age of Earth-Moon system

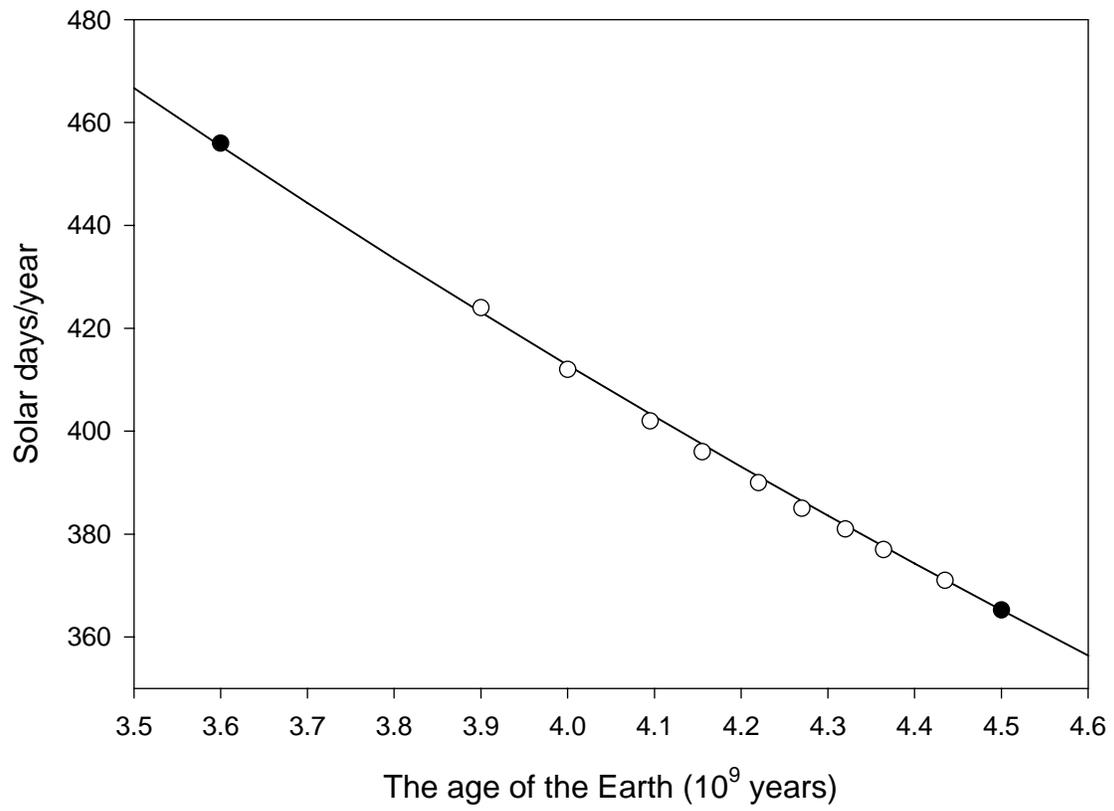

Figure 2, the solar days / year vs. the age of the Earth

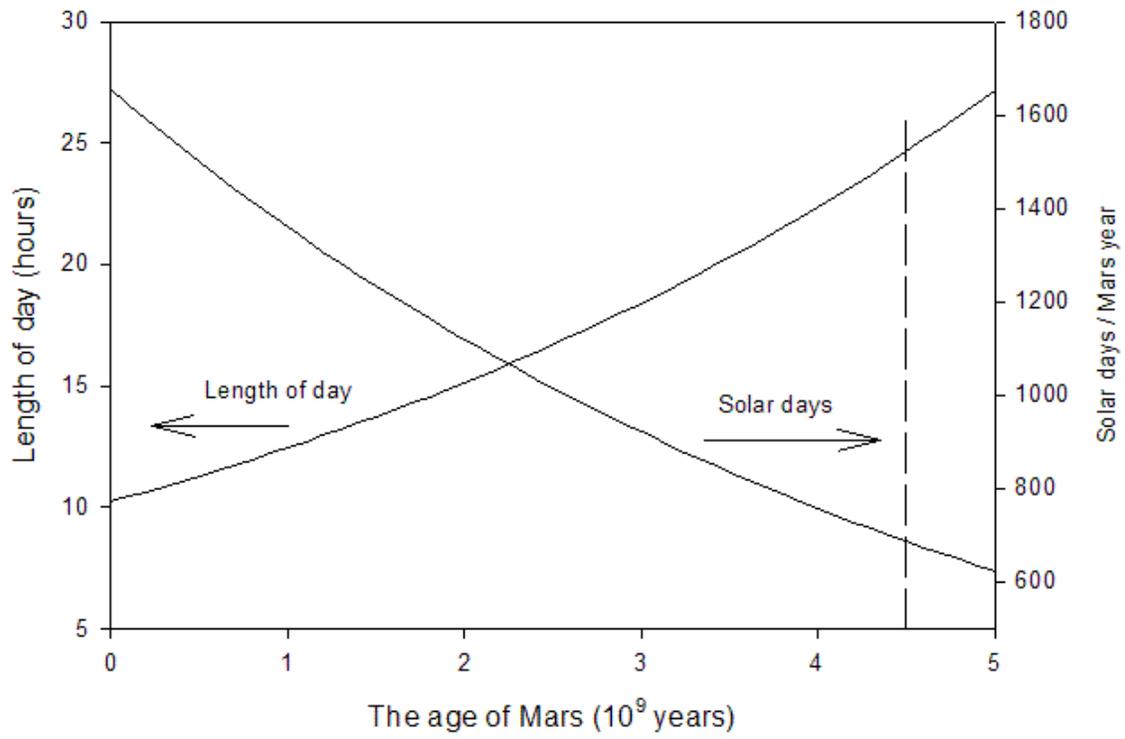

Figure 3, Mars's length of day and the solar days of year change with the age of Mars